# A Two-Stage Patient-Focused Study Design for Rare Disease Controlled Trials

Jian Yong[*], Sohaib H. Mohammad, Yan Yuan

**Affiliation:**   School of Public Health, University of Alberta, Edmonton, Alberta,
T6G 1C9, Canada

**\*Corresponding author**
Jian Yong, MSc
SOHU Fuxing Plaza, Tower A, 19thFloor
388 Madang Road Xintiandi Shanghai, China 200025
Email: yong3@ualberta.ca

Abstract

Objective: To develop a study design for rare disease clinical trials (RDTs) that can efficiently evaluate treatments, promotes access to new treatments during treatment development, and optimizes healthcare resource utilization for future treatment allocation, development, and prioritization.

Study design and setting: Comprehensive literature review and focus group discussion were conducted. We incorporated aspects other than treatment effect evaluation into the study design to address the multifaceted challenges facing RDTs.

Results: The four key considerations for RDTs are: 1) patients' opportunity to access the new treatment; 2) assessment of outcomes where clinically validated outcomes may be lacking; 3) patient heterogeneity; and 4) duration of the study and number of patients required. Our proposed study design has two stages. Stage 1 distinguishes patients who respond to the treatment from those who do not respond to the treatment after assigning them all to the experimental treatment. Stage 2 evaluates the treatment effect comparatively among patients responded in Stage 1.

Conclusion: In addition to treatment effect evaluation, RDTs can potentially great benefit rare disease patients and clinical practice by increasing opportunities to access experimental treatments and by providing relevant information that can be used for tailoring treatments to certain subgroups, aiding future research in treatment development, and improving healthcare resource utilization.

**What is new?**

This proposed study design has two stages: in Stage 1, all patients receive the experimental treatment; and in Stage 2, treatment effect is evaluated on patients whose outcome(s) improved in Stage 1 using either a cross-over design, series of n-of-1 trials, or a response-adaptive design, depending on the characteristics of the disease, treatments, and outcome(s) of interest. The design of both stages allows greater opportunity for patients to access the experimental treatment compared to a standard parallel design, which is often a critical consideration for patients.

This study design includes two salient features: (1) Responses to the experimental treatment are assessed at stage 1 with respect to patient characteristics to allow for better understanding of the experimental treatment responses, which is often inadequate for rare diseases. This also allows more tailored administration of treatment to future patients; and 2) there are three study design options for treatment effect evaluation in Stage 2, all of which allow greater opportunity for patients to access the experimental treatment.

# 1. Introduction

A disease is defined as rare if it affects less than 1 in 2,000 people in the population [1]. With over 6,000 diseases falling into this characterization, approximately 6-8% of people worldwide are afflicted with a rare disease [2-3]. Although rare diseases present a substantial burden to patients and healthcare systems, there are often little or no effective treatments available and the diseases themselves are poorly understood [4-6].

There are many challenges with using the standard randomized controlled trial (RCT) framework to evaluate treatment effect for rare diseases. These include small patient numbers, lack of validated clinical endpoints, lack of knowledge of the disease, and heterogeneity in patient characteristics, all due to the rarity of diseases [7,8]. High research and development costs for a small patient population present an additional challenge faced for orphan drug development, making them both scarce and expensive [7,9]. In turn, this limits the accessibility and availability of new treatments to patients[10,11], a critical issue which needs to be considered for rare disease patients and the trials of orphan drugs for them. A patient-centered trial design that promotes access to new treatments and optimizes healthcare resource utilization for future treatment allocation, development, and prioritization is urgently needed for rare diseases.

We propose a study design for rare disease trials (RDTs) that evaluates treatment effect comparatively under the standard trial requirements (i.e., randomization, blinding, allocation concealment), while promoting access to new experimental treatments for rare disease patients, and the understanding of the disease itself. Under this design, all enrolled patients have access to the experimental treatment; patient characteristics that are found to be associated with treatment response can inform tailoring of treatments based on patient characteristics in clinical practice.

## 2. Methods

The development of our study design followed the research plan outlined as follows:

(1) A comprehensive review of relevant literature to understand the issues surrounding the evaluation of rare disease treatments in clinical trials and to identify study designs that have already been applied and/or proposed.

(2) Discussions of key considerations for a study design among patients, researchers, clinicians, policy makers, and other stakeholders in order to address issues that rare disease clinical trials have in evaluating treatment effect.

(3) The development of a new study design for rare disease clinical trials, including key considerations that were developed as a result of the undertakings in 1) and 2).

We systematically searched in PubMed for English-language publications using multiple combinations of search terms (Table 1) in PubMed and included relevant literature published up to 31 May 2016. Based on the search results, JY scanned the titles and abstracts of 3,292 papers using three broad inclusion/exclusion criteria: 1) discussion of challenges with RDTs, 2) review of previously proposed RDT study designs or 3) application of RDT study designs. JY deemed 216 articles to meet these preliminary criteria and SM independently verified that all 216 articles were relevant given the study objective.

A more comprehensive review of all 216 papers was carried out by JY and SM and they mutually agreed to include 68 papers for a more comprehensive review (15 focused on challenges with RDTs, 17 reviewed previously proposed RDT study designs and 36 discussed applications of RDT study designs).

A list of key considerations was created in a small-group consultation setting with experts and annual meetings for the **P**romoting **R**are Disease Innovation through **S**ustainable **M**echanisms (PRISM) workgroup in 2013 and 2014. Together with the literature identified above, a list of important considerations for RDT's was developed.

Following additional discussions with two experts (Dr. Robin Casey, a physician involved in providing care for rare disease patients, and Dr. Robyn Lim, a senior scientific advisor at the Office of Legislative and Regulatory Modernization at Health Canada), a list of four key considerations was finalized. With the inclusion of these key considerations, we developed a new study design for RDTs.

## 3. Results

We first list the four key considerations. Next we describe our study design for RDT's where the target outcome of the disease is reversible[a]. In section 3.3, possible modifications that can be made to the proposed study design for non-reversible outcomes are described. Details on how this study design satisfies the four key considerations are given in the discussion section.

---

a.  Reversible outcomes refer to outcomes that can be deemed to be reversed back after a washout period to their baseline state. Non-reversible outcome are outcomes that cannot be reversed back to the baseline state of disease through washout.

## 3.1 List of considerations

The four key considerations for RDTs which were used to develop our study design are: 1) patients' opportunity to access the new treatment; 2) assessment of outcomes where clinically validated outcomes may be lacking; 3) patient heterogeneity; and 4) duration of the study and number of patients required.

In addition, we encourage investigators to include patients in the clinical trial committee prior to the onset of the trial. Several studies has shown that the inclusion of patients as informed decision makers, rather than just study subjects, can identify blind-spots that may have been missed by investigators [12, 13]. For example, the Outcome Measures in Rheumatology (OMERACT) group discovered outcomes which were important to patients with rheumatoid arthritis but had previously been missed by investigators, such as fatigue, foot pain, and sleep disturbances [14]. Thus, consulting or including patient representatives when planning of clinical trial can potentially elucidate new outcomes of interest and also identify how they may be measured during the trial.

## 3.2 Overview of the Proposed Study Design

In order to satisfy all four criteria, we propose a study design that has two stages. Stage 1 distinguishes patients who respond to the treatment ("responders") from those who do not respond to the treatment ("non-responders") after assigning them all to the experimental treatment. Stage 2 evaluates the treatment effect comparatively among those patients characterized as responders in Stage 1 (Figure 1).

### 3.2.1 Proposed Stage 1

Upon enrolment in Stage 1, detailed baseline characteristics of all enrolled study patients, which include demographic information (e.g., age and sex), specific clinical/biological information such as characterizations of the disease, symptoms, known biomarkers for the disease, and other physiological measurements, are collected. All enrolled patients are then given access to the experimental treatment. Patients' symptoms and clinical characteristics of the disease are monitored throughout Stage 1.

At the end of Stage 1, patients are characterized as either responders or non-responders. Responders are patients who respond positively to the treatment from patient self-reports and/or clinical evaluation that meet pre-specified criteria. If few or no patients are characterized as responsive at the end of Stage 1, the trial will be stopped with the conclusion that the treatment is overall ineffective. If responders exceed a pre-specified percentage, responders will proceed to Stage 2 while non-responders will be withdrawn from the trial. In the study protocol, investigators should clearly specify the criteria of responding to treatment and the minimum proportion of respondents that is needed to be observed at the end of Stage 1 to justify the proceeding to Stage 2.

Information about patient characteristics collected at baseline will be analyzed to assess the association between baseline patient characteristics and treatment response in order to gain understanding of the responses to the experimental treatment (e.g. side effects, quality of life improvement). This analysis can be done using standard statistical methods of binary outcomes (response vs. non response) such as logistic regression. Using the results from this analysis, patient subgroups can be created based on specific sets of characteristics that predict patients who are likely to benefit from and/or be harmed by the experimental treatment. This is useful information that can aid future treatment allocation, development, and prioritization since treatment accessibility and availability in rare disease populations are usually limited.

### 3.2.2 Proposed Stage 2 Framework

Following a washout period, patients who were characterized as responders in Stage 1 proceed to Stage 2 to evaluate the efficacy of the experimental treatment using a randomized controlled trial design. Depending on the characteristic of the outcomes of the rare disease under study, the investigator will choose an appropriate study design for evaluating the treatment effect in Stage 2.

Out of at least 14 different study designs [15-17], we shortlist three for use in Stage 2: cross-over design; series of n-of-1 trials design; and response-adaptive design (Table 2). These study designs were selected because they include randomization, allocation concealment, blinding of treatment allocation, and also allow for interim analysis. All three study designs further provide enhanced opportunities for enrolled patients to access the experimental treatment.

The use of interim analysis can potentially make treatment effect evaluation at Stage 2 more efficient. During Stage 2 interim analysis, if patients' outcomes have already been found to have substantially improved, then fewer patients and shorter study period than anticipated may be required for Stage 2.

For specific RDT scenarios, one of these study designs may be more suitable than others.  A summary of these three designs' features are given here:

1)  Cross-over design

- Definition: patients are randomly assigned to receive two or more treatments sequentially with wash-out periods between consecutive treatments and each patient acts as his/her own control [18].

- Assumptions: there is no treatment-period interaction and negligible carryover effect. Under these assumptions, statistically-valid comparison of two or more treatments is possible. [18].

- Advantages: since the comparison of treatment effect is within the same patient, this study design can more precisely estimate treatment effect than adaptive design where patient characteristics are heterogeneous [16,19-24].

- Limitations: 1) longer study duration is required than adaptive design because each patient receives more than one treatment; 2) sufficient washout period is required between each treatment received; and 3) consequences of dropouts are greater than adaptive design because more information is lost per dropout [18-24].

- Analysis: generalized linear model can be used in which the outcome can be binary or continuous. The effect of treatments, carry over, and periods are captured appropriately and evaluated in the model [25].

2) Series of n-of-1 trials design

- Definition: two or more treatments are consecutively and repeatedly given in a random order to each patient who contributes information to one n-of-1 trial. A series of n-of-1 trials are analyzed jointly for treatment effect evaluation [18, 20, 26].

- Assumptions: there is no treatment-period interaction and no carry over effect [18, 20, 26], under which comparing multiple treatments is statistically valid.

- Advantages: since this design is capable of assessing the effect of the treatment on each patient, it can estimate treatment effect more precisely when the disease is extremely rare and when patient characteristics are highly heterogeneous [18 20, 26].

- Limitations: 1) longer study duration is required due to a greater number of treatments received and/or more repetition of same treatments received than cross-over design and adaptive design; 2) similar to cross-over design, sufficient washout period is required between treatments received; and 3) the consequences of dropouts are greater than adaptive design because more information is lost per dropout.

- Analysis: meta-analysis can be performed for this study design by aggregating the data of each n-of-1 trial to obtain the average treatment effect for the population and individual patient, as well as estimates of between-patient variation [27, 28].

3) Response-adaptive study design

- Definition: a study design that allows modification of randomization schemes during the trial based on interim trial results. This is done by varying the probabilities of treatment assignment to increase the likelihood of patients being assigned to the superior treatment and minimizing the number of patients exposed to the inferior treatment [14, 18, 20, 29-34].

- Advantages: 1) carry over effect is not an issue because patients receive only one treatment; 2) the study duration is generally shorter than cross over trials and series of n-of-1 trials since patients only receive one treatment; 3) response-adaptive design is a parallel group design if no adaptation is made during the trial; and 4) the outcome response of previous patients can guide trials to assign the better treatment with high probabilities to newly recruited patients.

- Limitations: 1) not suitable for heterogeneous populations; and 2) complex statistical analysis is usually required under frequentist statistical inference [14, 18, 20, 29-34].

- Analysis: maximum likelihood estimation with consideration of correlation and permutation test are suggested as the analysis method for this study design [30, 33].

Stage 2 analysis, as in Stage 1, includes the perspective of enrolled patients for treatment effect evaluation since patient feedback is important information in determining the experimental treatment's effectiveness: it describes aspects of patients' experiences with the treatment that can be only obtainable from its actual users (e.g. pain, dizziness).

### 3.3 Potential modification for non-reversible outcomes

When the disease outcome is non-reversible, modifications to the study design are necessary to adequately evaluate treatment effect. For example, patients may show improvements in Stage 1 that are non-reversible; the continued administration of the experimental treatment in Stage 2, however, may do little to change their outcome from the "new" Stage 2 baseline. This can make it much harder to measure the effect of experimental treatment versus placebo or other treatment in Stage 2.

Thus, for non-reversible outcomes, Stage 2 in our proposed study design should be modified by recruiting new patients directly into Stage 2 instead of using Stage 1 patients. Patient characteristics collected in Stage 1 will be used to examine what subgroups of patients did not respond or had an adverse response to the treatment. New recruitment will have the same inclusion criteria as Stage 1 but will exclude patients who possess characteristics that were found to be associated with treatment non-response in Stage 1. By excluding patients who are unlikely to benefit from the experimental treatment, Stage 2 will focus the treatment effect evaluation on patients who are more likely to respond.

## 4. Discussion

It is estimated that up to 8% of the world population has a rare medical condition [2]. Since rare disease treatments often cost much more than what many patients can afford, funding for the treatment access of a small population could significantly impact the budget of healthcare systems [2]. Furthermore, since there is limited information related to the epidemiology of many rare diseases and rare disease treatment effects, treatment in clinical practice might often be done with limited evidence , which may result in poor utilization of healthcare resources [1]. Thus, a study design that can aid prioritization of resource utilization in treatment development, treatment allocation in clinical practice and provides more treatment access opportunities to patients, as well as evaluates the treatment's effects, would make effective treatments more available for rare disease patients and would better utilize the resources of developing clinical trials and public funding subsidization. In sections 4.1-4.4, we discuss why each key consideration is important in these regards and then describe how each stage of our study design satisfies the consideration. A summary is given in Table 3.

**4.1 Consideration 1: Patient Opportunity to Receive the Experimental Treatment**

For many rare diseases, there may be a lack of industry incentive to develop treatments for several reasons: 1) there may be insufficient scientific interests; 2) a small number of patients to treat is perceived as relatively low return on investment which may impede treatment development; and 3) the poorly understood etiology means that treatment development is less certain due to factors such as a lack of pharmacological targets for intervention. As an ethical consideration, it would greatly aid rare disease patients if RDTs made treatment access a priority given the scarcity of treatment access opportunities in the rare disease community.

In order to enhance treatment access for patients in the trial, Stage 1 of our proposed study design is an "enrichment" stage where all patients enrolled in the trial receive the experimental treatment. In addition, Stage 2 for treatment effect evaluation employs study designs that give patients more opportunity to access the experimental treatment compared to a parallel group design. Specifically, n-of-1 and cross-over study designs ensure that all patients receive the experimental treatment during the treatment effect evaluation in Stage 2. For response-adaptive designs, the probability of receiving a superior treatment increases as the trial progresses, so patients have a greater chance of being selected into the superior treatment arm.

## 4.2 Consideration 2: Assessment of outcomes where clinically validated outcomes may be lacking

The effectiveness of a treatment in clinical trials is usually based on well-characterized clinical outcomes [5]. If validated clinical outcomes are lacking, then it is challenging to justify the treatments effect on patients' health conditions. Therefore, it would aid treatment effect evaluation if RDTs can evaluate treatment effects where no clinically validated endpoints for a disease are available.

In section 3.1, we suggest including patients as shared decision makers, prior to the start of the trial. Our study design itself also takes outcome measurement into consideration in both stages. In Stage 1, we measure a range of modifiable patient characteristics before and after the treatment is administered and monitor the changes. This information can then be incorporated in Stage 2 to define endpoints. In both Stages 1 and 2, patient-feedback is considered in the evaluation of treatment response. The measurements of modifiable patient characteristics and patient-feedback provide information about outcomes/endpoints relevant to evaluate treatment effect: they may also provide information about how the administration of the experimental treatment can be developed to be more patient-friendly [4].

## 4.3 Consideration 3: Patient Heterogeneity

Many rare diseases populations are known to be heterogeneous [2]. Patients with certain characteristics might respond to the experimental treatment whereas others might not. Since rare disease treatments are often expensive and scarce, the association between patient characteristics and treatment response is an important consideration for treatment development and prioritizing treatments for patients who are likely to respond in the clinical practice. This could also provide future direction for treatment development by potentially providing information regarding plausible biological mechanisms of treatments being more effective for certain patients than others.

To account for patient heterogeneity in Stage 1, we screen for responders and explore the patient characteristics associated with treatment response. This information is then used to create patient subgroups based on patient characteristics. For example, if a common biological marker exists among all responders but not among non-responders, this can inform new areas for pharmaceutical intervention and can also prioritize treatment allocation in the clinical setting. In Stage 2, only patients who responded to the treatment in Stage 1 are assessed, and hence the evaluated treatment effect is more specific to certain groups of the patient population that may benefit from the treatment.

## 4.4 Consideration 4: Duration of the study and recruitment of a sufficient number of patients

An efficient study design can get an effective intervention to rare disease patients more quickly by avoiding an unnecessarily long study period. For RDTs, a long study period is often required to recruit a sufficient number of patients into the study: finding/identifying the specific rare disease patients who may be geographically dispersed takes time. For these reasons, investigators should consider how the trial can be made more efficient without sacrificing patient safety and the integrity of the trial.

In our study design, investigators are asked to specify, in the study protocol, the minimum proportion of patients responding to treatment in Stage 1 to justify proceeding to Stage 2. If this proportion of responding is not met in Stage 1, the trial is stopped without proceeding to Stage 2. For Stage 2, interim analysis is used; again, if sufficient response is not seen, as specified in the study protocol, the trial is stopped. This improves efficiency and also helps with resource utilization by stopping the trial, should the treatment be found ineffective at an earlier point.

Additionally, in section 3.5, we describe a modification that can be made for non-reversible outcome. However, this modification can also be used to enhance efficiency. If a clear pattern of patient characteristics are found to be associated with treatment response in Stage 1, the study design can be modified by having newly recruited patients enrol directly into Stage 2. One limitation of this modification is that if a response-adaptive design is used in Stage 2, some patients may not get the opportunity to access the experimental treatment. The impact of such a loss of opportunity should be considered prior to making such a study design modification.

## 4.5 Comparison with randomized withdrawal design

Our study design shares some similarities with a previously proposed RDT study design, the randomized withdrawal design [34-42]. Specifically, both our proposed study design and the randomized withdrawal design have 2 phases/stages; there is an initial enrichment stage where all recruited patients receive the experimental treatment and only patients who respond to the treatment continue to the second stage for treatment effect evaluation.

The salient features of our design are: 1) the association of patient characteristics and responsiveness to the experimental treatment are analyzed at the end of Stage 1; and 2) there are three study design options to evaluate treatment effect in Stage 2. Regarding feature 1), we gain knowledge on patient subgroups for which the treatment is effective, which has important implications for treatment development, allocation, and prioritization. On the other hand, the randomized withdrawal design is often motivated to "enrich" the patient population and increase study power in the second stage's efficacy evaluation [37]. With feature 2), more study design options in Stage 2 makes it possible to evaluate treatment effect for a greater number of rare diseases with the proposed designs in our framework, such as cross-over design, series of n-of-1 trials, compare to randomized withdrawal design in which only the parallel group design is used in Stage 2. In addition, our design considers a greater scope of specific RDT issues that patients, healthcare providers, funders and treatment developers face as described above.

**4.6 Limitations**

This proposed RDT framework has some potential limitations which should be considered. In general, investigations of the association between patient characteristics and treatment response in a clinical trial are recommended to be exploratory: the association may be specific to some aspects of the trial (e.g., season). The results on the association from Stage 1 should, therefore, cautiously be generalized to the target population. Our suggestion to alleviate this limitation is to ensure sufficient sample size for identifying the associations in Stage 1 as well as soliciting from patients and care providers possible specific aspects of the trials that may potentially bias the trial results.

It is also possible that the treatment is found to be effective following Stage 2, but there is ambiguity about how the characteristics of responders and non-responders differ. In such a scenario, the trial has a limited capacity to prioritize treatment based on patient characteristics. However, the information about treatment effects is still valuable for stakeholders in determining the proportion of the target population that is expected to respond to the treatment. Given that rare disease treatments often are very expensive, information on the percentage responding and average effect size for responders will help with resource utilization.

Since the development of this study design did not consider regulatory aspects, our study design might be modified after the inclusion of regulatory aspects.

## 5. Conclusion

In addition to treatment effect evaluation, RDTs can potentially be of great benefit to rare disease patients and clinical practice by increasing opportunities to access experimental treatments and by providing relevant information that can be used for tailoring treatments to certain subgroups, aiding future research in treatment development, and improving healthcare resource utilization. Future work in applying our proposed design to rare disease clinical trials is needed in order to evaluate its robustness in practice. Although this design has some limitations and may not be suitable for all rare disease scenarios, it may serve as a basis for the future development of RDT study designs that aim to better fulfill the needs of the rare disease community.

## Acknowledgement

The research was supported by a team grant to the **P**romoting **R**are Disease Innovation through **S**ustainable **M**echanisms (PRISM) team from the Canadian Institutes of Health Research. The authors thank Dr. Robyn Lim from Health Canada for thoughtful discussions.

**Table 1. Search terms for the identification of articles on rare disease clinical trial topics**

| | | |
|---|---|---|
| • Rare disease | • Randomized withdrawal | • Crossover |
| • Research Design | • Adaptive design | • Series N-of-1 |
| • Epidemiologic Methods | • Response adaptive design | • Small Clinical Trial |
| • Clinical trials | • Sequential design | • Orphan Drug |
| • Ranking and Selection | • Enrichment design | • Orphan |
| • N-of-1 | • Clinical Trials as Topics | • Analysis |

**Table 2. Study design options for Stage 2 treatment effect evaluation**

### A) Cross-Over Design

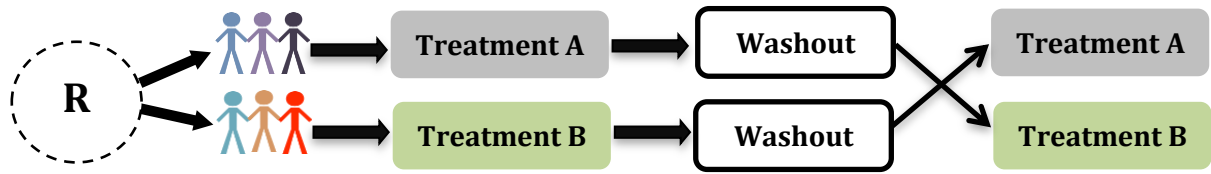

**Definition:**
Randomly assign sequences of two treatments to patient groups

**Advantages:**
- Comparison of two treatments
- No treatment-period interaction
- No carry over effect
- Heterogeneous populations

**Limitations:**
- Longer study duration
- Sufficient washout period needed
- Dropouts have greater influence

### B) Series of N-of-1 Trials

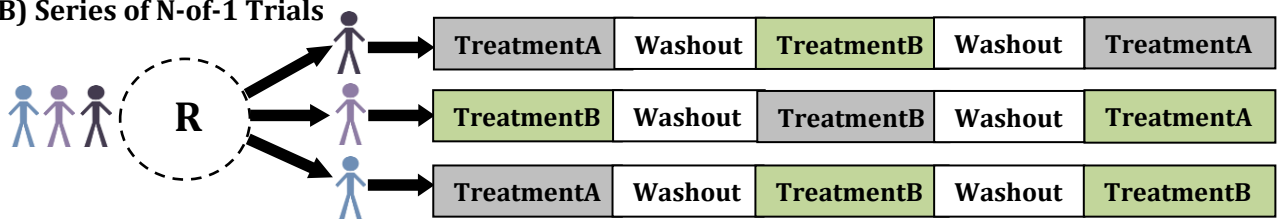

**Definition**
Randomly assign sequences of multiple treatments to only one patient in a trial for a series of trials

**Advantages:**
- Comparison of two or more treatments
- No treatment-period interaction
- No carry over effect
- Heterogenous populations
- Extremely rare disease

**Limitations:**
- Longer study duration
- Sufficient washout period needed
- Dropouts have greater influence

### C) Response Adaptive Design

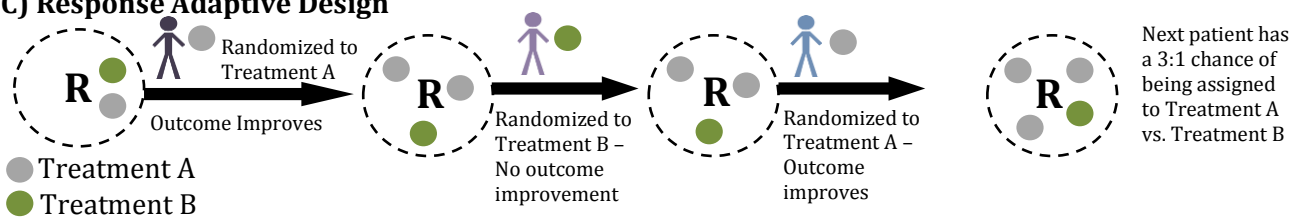

**Definition:**
Each treatment's chance of being assigned to patients is adapted based on accumulated patients' responses to treatments

**Advantages:**
- Evidence of one superior treatment
- Shorter study duration preferred

**Limitations:**
- Not suitable for heterogeneous populations
- Complex analysis

**Note:** "R" represents randomization. Treatment A and B refer to an experimental treatment and a standard treatment, respectively. For response-adaptive design, each patient receives Treatment A or B; the probability of receiving a specific treatment depends on the number of balls for that treatment in ths "ball-in-urn" design. The number of balls in the urn is for illustration due to limited space. The required number of balls varies upon studies.

**Table 3. Summary of how the Proposed Design (Stage 1 and 2) Satisfy the Four Key Considerations**

| Key Considerations | Rationale | Stage 1 | Stage 2 |
|---|---|---|---|
| **1. Patient opportunity to access the new treatment** | Due to a lack of accessible and effective treatments, treatment access is a high priority for rare disease patients | All patients receive the experimental treatment | Study designs give more opportunity to patients to receive treatment or the superior treatment than parallel group design does |
| **2. Assessment of outcomes where clinically validated outcomes may be lacking** | The trial should evaluate treatment effect on all rare disease patients even without validated clinical endpoints so as to benefit clinical practice | Patient-feedback on treatment effect | Patient feedback on treatment effect |
| **3. Patient heterogeneity** | Treatments should be prioritized to those patients most likely to benefit from them | - Patients are characterized into responders and non-responders<br>- Measurement of patient characteristics and association with treatment response | - Only patients who responded to the treatment in Stage 1 are assessed<br>- N-of-1 and cross-over study designs in Stage 2 use patients as their own controls |
| **4. Duration of the study and number of patients required** | An efficient study design will get an effective intervention to patients sooner and will aid resource utilization | If sufficient Stage 1 improvement is not seen, the trial is terminated | - Interim analysis is performed to justify continuing the trial<br>- Only responders from Stage 1 patients are evaluated for treatment effects |

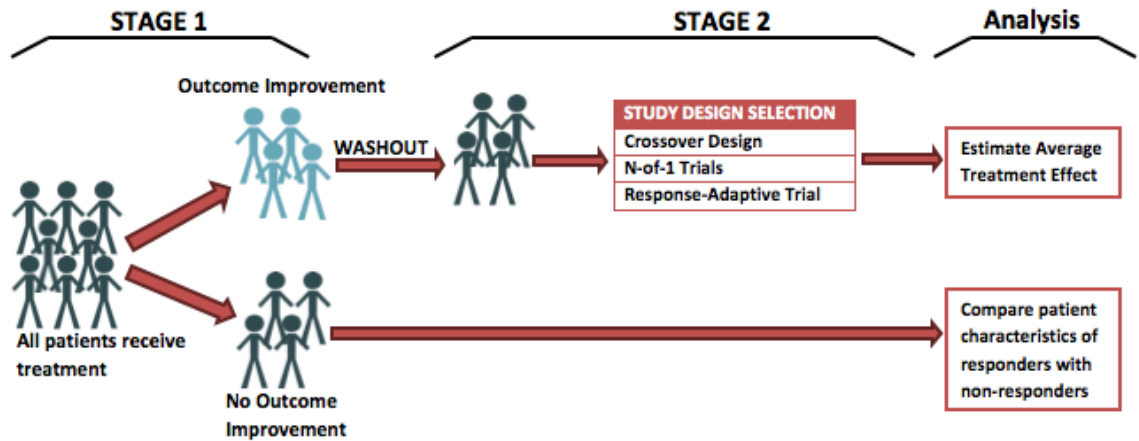

**Figure 1. Overview of the proposed framework**

During Stage 1, all patients receive the experimental treatment and are identified as "responders" or "non-responders" based on their outcome improvement. Following a washout period, responders proceed to Stage 2 for comparative evaluation of treatment effects among them. Analysis has two goals: 1) estimation of the average treatment effect on responsive patients; and 2) comparison of the characteristics of responsive and non-responsive patients.